\documentclass[twocolumn,prl,amsmath,amssymb]{revtex4-1}

\usepackage[pdftex]{graphicx}
\usepackage{dcolumn}
\usepackage{bm}
\usepackage[T1]{fontenc}
\usepackage[utf8]{inputenc}
\usepackage{natbib}
\usepackage{hyperref}
\usepackage{color}
\usepackage{url}

\usepackage{amssymb,amsmath,amsfonts,graphicx}
\usepackage{comment}
\usepackage{subfigure}
\usepackage{xcolor}
\usepackage[english,francais]{babel}
\begin{document}
\setcounter{totalnumber}{10}

\definecolor{RedPastel}{RGB}{255,125,82}
\newcommand{\red}{\color{RedPastel}}  

\title{Frozen impacted drop: \\ From fragmentation to hierarchical crack patterns}

\author{Elisabeth Ghabache}
 \affiliation{Sorbonne Universit\'es, UPMC Univ Paris 06, UMR 7190, Institut Jean Le Rond d'Alembert, F-75005, Paris, France\\
CNRS, UMR 7190, Institut Jean Le Rond d'Alembert, F-75005, Paris, France}
\author{Christophe Josserand}
 \affiliation{Sorbonne Universit\'es, UPMC Univ Paris 06, UMR 7190, Institut Jean Le Rond d'Alembert, F-75005, Paris, France\\
CNRS, UMR 7190, Institut Jean Le Rond d'Alembert, F-75005, Paris, France}
\author{Thomas S\'eon}
 \affiliation{Sorbonne Universit\'es, UPMC Univ Paris 06, UMR 7190, Institut Jean Le Rond d'Alembert, F-75005, Paris, France\\
CNRS, UMR 7190, Institut Jean Le Rond d'Alembert, F-75005, Paris, France}

\date{\today }

\begin{abstract} 
We investigate experimentally the quenching of a liquid pancake, obtained through the impact of a water drop on a cold solid substrate ($0$ to $-60^\circ$C). We show that, below a certain substrate temperature, fractures appear on the frozen pancake and the crack patterns change from a 2D fragmentation regime to a hierarchical fracture regime as the thermal shock is stronger. The different regimes are discussed and the transition temperatures are estimated through classical fracture scaling arguments. Finally, a phase diagram presents how these regimes can be controlled by the drop impact parameters. 
\end{abstract}

\maketitle


When molten glass drips into cold water, the outside cools - and shrinks - faster than the inside, creating pent-up tension in the so-called  Prince Rupert's drop, known since before $1625$ to have very striking mechanical properties~\cite{Merrett1662,Chandra94}. Indeed, while the drop's head stays impervious to even the strongest blows, flick the tail and the whole drop shatters in a myriad of small pieces, in less than a millisecond.
In the same way, fragmentation is in fact present in many physical processes, from jet atomization to bubble bursting in 
fluids~\cite{Marmo04,ARViller07,Lhuissier12}, from spaghetti breaking~\cite{Audoly05} to popping 
balloons~\cite{Moulinet15} or broken windows in solids~\cite{Vandenberghe13a,Vandenberghe13b}.
It is related to diverse applications such as comminution~\cite{Comm15}, shell case bursting~
\cite{Mott47}, ash generation during eruption~\cite{Kokelaar1986, Liu2015} or meteoric cratering~\cite{Sagy04} for instance. 

Fragmentation is thus a sudden process, where the considered domain mainly divides in one go, with a very fast cracks front propagation. At least as ubiquitous, there exists a complete different crack morphology where space-dividing pattern shows a strong hierarchy of slower fractures~\cite{Bohn2005}. Fractures develop successively, and each new fracture joins older fractures at a typical angle close to ninety degrees~\cite{Shorlin2000, Lazarus2011}. Such patterns are usually observed when the shrinking of a material layer is frustrated by its deposition on a non shrinking substrate, such as drying-induced cracks in mud \cite{Kindle1917, Korvin1989}, coffee \cite{Groisman1994}, colloidal silicas \cite{Pauchard1999}, industrial coating~\cite{Xu2009} or artistic painting~\cite{Pauchard2007}. 

In this paper, we investigate experimentally the quenching of a liquid pancake that is obtained through the impact of a water drop on a very cold solid substrate. We show for the first time that, as a function of the substrate temperature, the crack patterns produced by the thermal shock, change from a  2D fragmentation regime to a hierarchical fracture regime, and 
the transition temperatures are estimated and discussed.

\begin{figure}[h]
\center
\includegraphics[width=1.\linewidth]{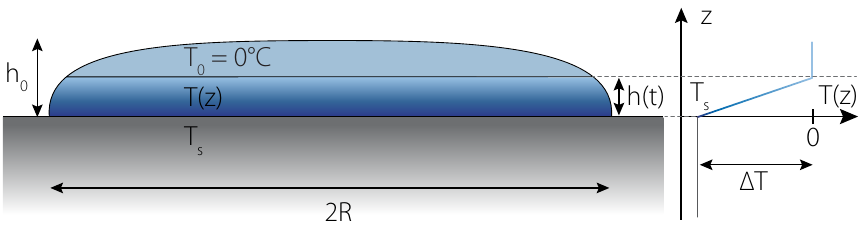}
\caption{Scheme of the frozen pancake obtained after a liquid drop impacted a cold substrate. The pancake has a radius $R$ and a typical thickness $h_0$. The substrate, at a temperature $T_s$, cools the pancake, so that a layer of thickness $h(t)$ is frozen, above which the liquid is at freezing temperature $T_0$.}
\label{Schema_goutte}
\end{figure}

The experimental setup consists in dropping a drop of water, with a diameter $D_0=3.9$ mm, on a steel substrate, so as to form a liquid pancake of radius R and typical thickness $h_0$ (see Fig.~\ref{Schema_goutte}). Under this simplified geometry $h_0$ can be estimated by balancing the volume of the drop with that of the cylindrical pancake  $h_0 = D_0^3/6R^2$.
The impact velocity is close to the free fall one: $U_0 \sim \sqrt{gH}$ where $H$ is the falling height. 
Throughout most of the paper, falling height will be kept constant at $H = 36$ cm. Subsequent pancake radius is : $R \simeq 8$ mm, from which pancake thickness  can be deduced : $h_0 \simeq150 \mu$m.
The temperature of the substrate $T_s$ is typically varied from the water freezing temperature, $0^\circ$C, to  $-60^\circ$C. It is reached by plunging a large cube of  stainless steel ($10^3$ cm$^3$) into a liquid nitrogen bath until the desired temperature is reached. The whole experiment is made into a controlled atmosphere box in order to avoid frost formation on the substrate. 
Because of the small experiment time (max. $\sim 1$ s), we can consider that the substrate remains at constant temperature during the dynamics. 
The drop dynamics is visualized using a high-speed camera recording the spreading from the top.

\begin{figure*}[th]
\center
\includegraphics[width=1.\linewidth]{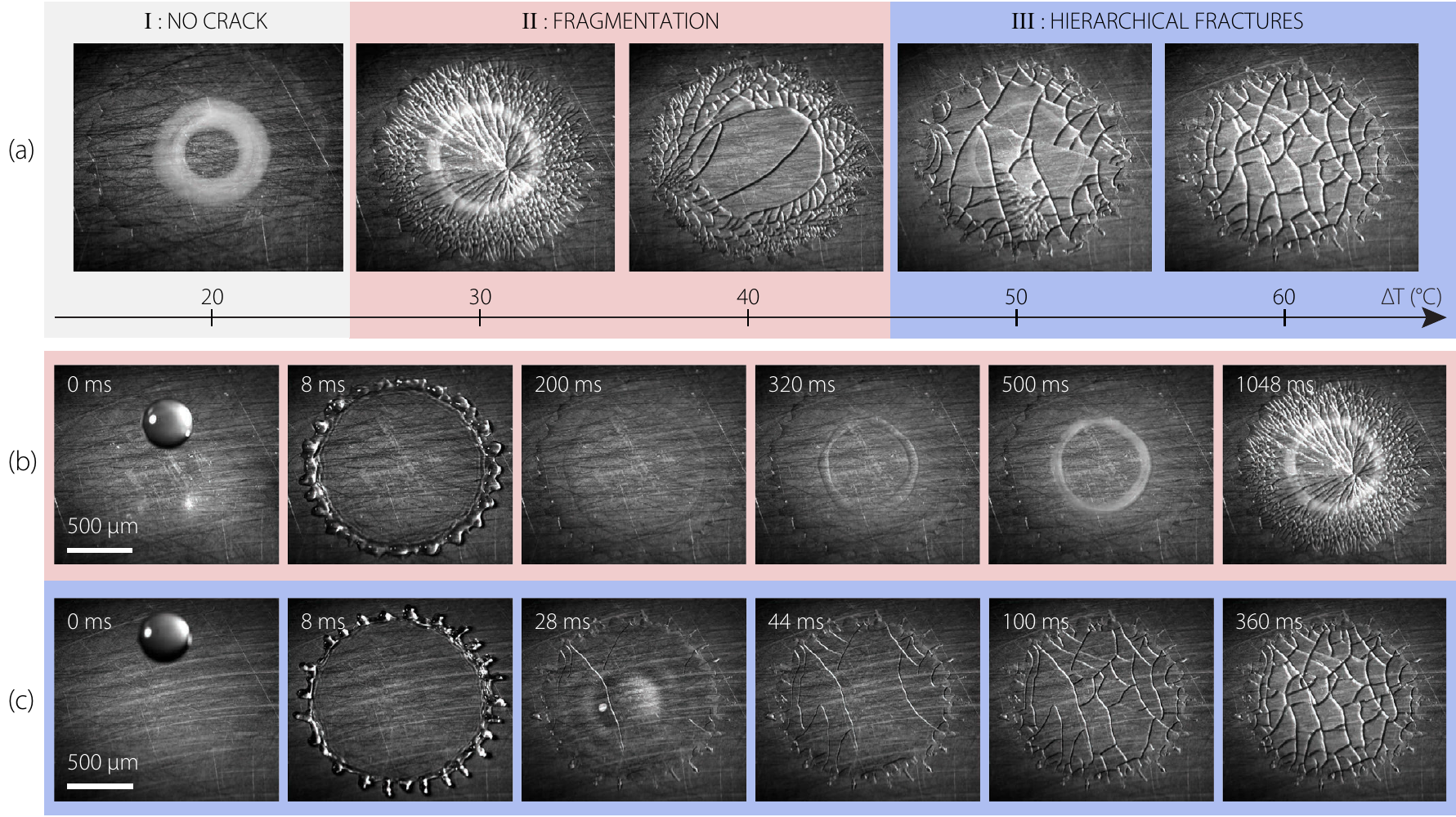}
\caption{ (a) Frieze presenting snapshots of the frozen pancakes formed after a water drop impacted, from a falling height $H =  36$ cm, a cold substrate at various temperature $T_s = -20.0^\circ$, $-31.1^\circ$, $-41.2^\circ$, $-50.3^\circ$ and $-59.6^\circ$C from left to right (with $\Delta T=-T_s$). 
Depending on $\Delta T$ the frozen pancake presents different crack patterns that can be gathered into three different regimes : I - no cracks, II - fragmentation regime, III - hierarchical fractures regime.  The transition temperatures are : $\Delta T_\text{I-II}^{\text{(exp)}}\sim27^\circ$C and $\Delta T_\text{II-III}^{\text{(exp)}}\sim42^\circ$C.
(b) Sequence showing the drop impact and solidification dynamics preceding the fracture pattern observed on the second image of the frieze (a) : $T_s=-31.1^\circ$C. 
(c) Sequence preceding the fracture pattern observed on the fifth image of the frieze (a) : $T_s=-59.6^\circ$C. 
On these two sequences, the time and the scale bar are on the images.
}

\label{cracks}
\end{figure*}


Figure~\ref{cracks} (b) and (c) present these time sequences, each one for a different substrate temperature, respectively -31$^\circ$C and -60$^\circ$C. In both impact sequences, the drop spreads on the substrate until it reaches its maximum diameter, captured on the second image. 
During this phase, the droplet remains liquid but a thin layer of ice forms upon contact with the substrate. 
In contrast with the situation at room temperature \cite{JTAR16}, almost no retraction of the drop is further observed since it is pinned on the solid substrate, most probably by this ice layer. 
Instead, capillary waves propagate on the spread droplet that has now the shape of a pancake. In the mean time, the solidification of the drop occurs, observed on Fig.~\ref{cracks} (b) through a front that develops radially from the pancake edge towards its center, forming eventually a donuts that solidifies (t $\sim$ 500ms), and that is due to the complex dynamics of the solidification front~\cite{Brunet14}. 
After that point (t $\sim$ 500ms on Fig.~\ref{cracks} (b)), the whole pancake is frozen and keeps cooling. It therefore shrinks more and more, but the adhesion to the solid substrate limits this ice contraction. This frustration causes mechanical tensions that are suddenly relaxed by the formation of a pattern of fractures. This remarkable dynamics, called fragmentation, is a 2D equivalent to the Prince Rupert's drops shattering, described in the introduction.
This solid fragmentation seems to propagate radially from a nucleation point. Experimental estimation gives a high front propagation velocity, typically between $800$ and $1000$ m.s$^{-1}$, which is a fraction of the Rayleigh wave speed. 

Figure~\ref{cracks} (c) presents the same drop impact experiment but on even colder substrate (-46$^\circ$C). In this case, shortly after drop has pinned, while ripples are still visible, first fractures are observed on a growing ice layer (t $\sim$ 28 ms). Then more cracks propagate, hierarchically, by successive division of the frozen drop.
The crack pattern is here typical of hierarchical fractures~\cite{Bohn2005}, with younger crack joining the older one at an angle close to 90$^\circ$. The domains are larger and consequently less numerous than in the fragmentation regime. Note that, if this particular cracking dynamics is very similar to what is observed in the case of desiccation~\cite{Groisman1994}, here the time scales are much shorter. 


To summarize the qualitative description of our experiment, the main different patterns are shown on Fig.~\ref{cracks} (a) as a function of the temperature difference $\Delta T=T_0-T_s$, where $T_s$ is the substrate temperature and $T_0=0^\circ$C is the water freezing temperature. They are gathered in three different regimes :
\begin{itemize}
\renewcommand{\labelitemi}{$\bullet$}
\item I : at low $\Delta T$, the solid pancake remains smooth, no cracks are present. 
\item II : the fragmentation regime, at intermediate $\Delta T$, the cracks appear suddenly from a nucleation point. 
\item III : the hierarchical regime, at high $\Delta T$, the cracks  are formed step by step in successive sequence. 
\end{itemize}
The two sequences described above, Fig.~\ref{cracks} (b) and (c), belong respectively to the beginning of regime II and the end of regime III, where the dynamics is net. We observe that, close to the transition between the regimes, intermediate cases appear, fragmentation only on the edge of the pancake or mix between fragmentation and hierarchical fractures. 
It is also worth emphasizing that while the fragmentation occurs after the whole pancake has solidified, the hierarchical cracks are usually formed during the solidification phase: if the bottom part of the pancake is solid, a liquid layer is still present on the top. 
Finally, this experiment is, to our knowledge, the first example where it is possible to pass continuously from a fragmentation to a hierarchical regime using a simple control parameter. Thermal shock in ceramic~\cite{Korneta1998, Lahlil2013} might have comparable behavior, but this has not been observed so far.



These different regimes can be understood using classical fracture arguments~\cite{Freund}: indeed, once the liquid pancake has solidified, the new solid is submitted to a rapid thermal contraction as substrate temperature is below 0$^\circ$C. 
If the ensuing deformation energy is high enough, fractures can appear in the frozen pancake. This mechanisms can be quantified using energy balance arguments~\cite{Mokhtar,Marigo}. : we assume a linear isotropic elastic behavior of ice, with a Young's modulus $E=9.33$~GPa. 
Its thermal contraction induces a deformation tensor field $ {\bf \varepsilon}_{th}({\bf x},t)=\alpha \delta T {\bf I}$ where $\alpha=5.3 \cdot 10^{-5} \, {\rm K}^{-1}$ is the ice thermal expansion coefficient~\cite{Petrenko1999}, ${\bf I}$ 
the identity tensor and $ \delta T=T_0 - T({\bf x},t)$, with $T({\bf x},t)$ the local time-dependent 
temperature in the ice domain. The density of elastic energy induced by the thermal contraction reads therefore:

\begin{equation} 
{\cal E} =\frac12 E {\bf \varepsilon}_{th}: {\bf \varepsilon}_{th}=\frac32 E \alpha^2 \delta T^2.
\label{DensityEnergy}
\end{equation}
A fracture in a brittle material consists in the formation of a new interface, associated to an energy per unit surface, the so-called Griffith energy, $G_c \sim 1~{\rm kg} \cdot {\rm s}^{-2}$ for the ice.
Balancing the elastic energy due to the thermal contraction of a cubic ice of length $L_c$, with homogeneous temperature $T_s$, $3E (\alpha \Delta T)^2 L_c^3/2$, with the energy of a crack breaking the cube in two part $2G_c L_c^2$,  leads to the introduction of the Griffith length : 
\begin{equation}
 L_c= \frac{4G_c}{3E \alpha^2 \Delta T^2}.
 \label{Griffith_length}
\end{equation}
Above this typical length, breaking the shrinking solid becomes energetically favorable.

In our system, two regimes can therefore be identified in the crack formation, depending on the ratio between the Griffith length $L_c$ and the typical height $h_0$ of the liquid pancake. If $h_0 \gg L_c$ one expects that the cracks appear before the whole solidification of the pancake, when a solid ice layer of thickness of the order of $L_c$ is formed. This is the behaviour observed on Fig.~\ref{cracks} (c) and therefore corresponding to regime III. On the other hand, for $h_0 \ll L_c$, the cracks would appear only after the total solidification of the pancake. We identify this latter case with the regime II where the frozen pancake fragments into a myriad of small pieces of typical size $h_0$~\cite{Bohn2005}.

In the aim of estimating the appearance temperature of first cracks at the frontier between regime I and II , $\Delta T_\text{I-II}$, let us consider the case $h_0 \ll L_c$. The energy balance imposes, that the total elastic energy in the frozen pancake is greater than the surface energy of all the fractures, namely:
$$ \frac{3}{2} E \alpha^2 \Delta T^2 \pi R^2 h_0 \ge 4 \frac{\pi R^2}{h_0^2} G_c h_0^2,$$
where the ratio $\pi R^2/h_0^2$ is the number of pieces of typical size $h_0$ formed by the fragmentation. It leads to the relation:
\begin{equation} 
\Delta T^2 \ge \Delta T_\text{I-II}^2= \frac{8 G_c}{3E \alpha^2 h_0}. 
\label{DeltaT_I-II}
\end{equation}
With, for $\Delta T > \Delta T_\text{I-II}$ cracks are energetically favorable while for $\Delta T < \Delta T_\text{I-II}$ no cracks should be observed. Taking the values of E, $G_c$ and $h_0$ given above, leads to $\Delta T_\text{I-II} \sim 26^\circ$C, which is in amazingly good agreement with the experimental transition temperature to fragmentation $\Delta T_\text{I-II}^{\text{(exp)}}\sim27^\circ$ (Fig.~\ref{cracks}). 

On the other hand, when $h_0 \gg L_c$, fractures can form before the full solidification of the liquid pancake and we identify there the regime III, where the cracks appear step by step. In this case, the solid layer of thickness $h(t)$ grows with time as the pancake freezes (see Fig.~\ref{Schema_goutte}). The diffusive heat flux through this solid layer, $Q= -\lambda \partial_z T$, is then balanced, at the solidification front, by the solidification rate $-\rho_s L \dot{h}(t)$. Here, $L=333.5~{\rm kJ}\cdot {\rm kg}^{-1}$ is the ice-water latent heat per unit mass, $\rho_s=920~{\rm kg}\cdot{\rm m}^{-3}$ the density of ice and $\lambda=2.4~{\rm W}\cdot{\rm m}^{-1}\cdot{\rm K}^{-1}$  its thermal conductivity~\cite{Petrenko1999}.
This gives a time scale for the solidification process, $ \tau_s =\frac{\rho_s L h^2}{\lambda \Delta T}$. Comparing this latter to the time scale of heat diffusion $ \tau_d =\frac{h^2}{D}$ leads to the Stefan number :
$${\rm St}=\frac{C_p  \Delta T}{L}=\frac{ \tau_d}{\tau_s},$$
where, $D=\frac{\lambda}{\rho_s C_p}=1.3 \cdot 10^{-6} ~{\rm m}^2\cdot {\rm s}^{-1}$ is the heat diffusion coefficient of the ice. 
In our experiments, the Stefan number is always smaller than one, indicating that the diffusion process is always faster than the solidification dynamics. 
Therefore, we can consider that the temperature field in the ice layer is in a quasi-stationary regime, obeying to the stationary diffusion equation. Taking a simple horizontal ice layer of heigh $h(t)$ it reads $ \partial_{zz} T=0$, with the boundary conditions $T(0,t)=T_s $ and $T(h(t),t)=T_0 $ since the temperature at the solidification front $z=h(t)$ is the fusion temperature. It leads to the linear temperature field:
\begin{equation}
T(z,t)= T_s + \Delta T \frac{z}{h(t)}. 
\label{TempProfil}
\end{equation}

Now the  time dependent diffusive heat flux through the ice can be computed $Q= -\lambda \partial_z T=-\lambda\Delta T/h(t)$,  and the balance with the solidification rate $-\rho_s L \dot{h}(t)$, gives the following time evolution for the ice layer 
$$h^2(t)=\frac{2 \lambda \Delta T}{\rho_s L} t= 2\text{St}D~t,$$
taking $h(0)=0$.
Considering that the formation of the first crack happens when $h(t_c) \propto L_c$ (Eq.~\ref{Griffith_length}), it gives for the time of 
cracks appearance in regime III:
\begin{equation}
t_c \propto \frac{8 \rho_s L G_c^2}{9 \lambda E^2 \alpha^4 \Delta T^5}.
\label{tf}
\end{equation} 

\begin{figure}[h]
\center
\includegraphics[width=1.\linewidth]{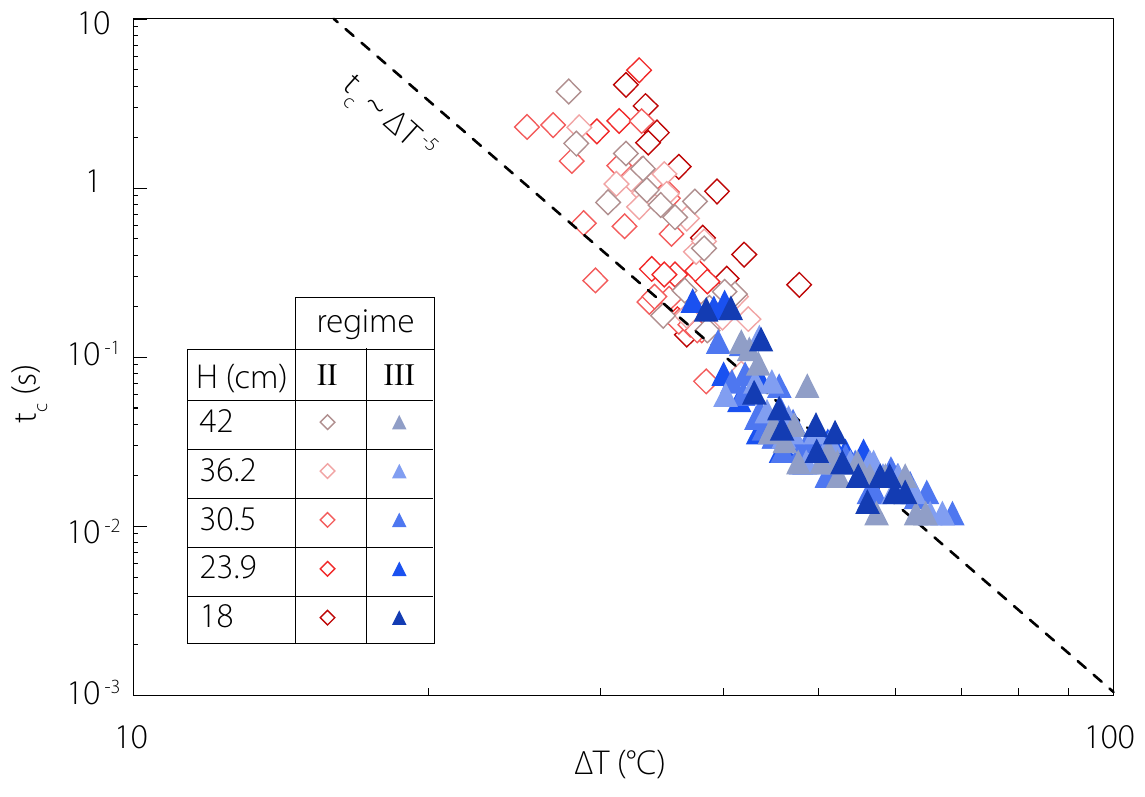}
\caption{Appearance time of the first crack, $t_c$, plotted as function of $\Delta T = T_0 - T_s = - T_s$ with $T_s$ the substrate temperature, for five different falling heights of the impacting drop. $t_c$ is determined considering initial time when drop reached its maximum spreading diameter after impact. The open diamonds correspond to the fragmentation regime (II) while the closed triangles correspond to the hierarchical fracture regime (III). The dashed line, representing $t_c \propto \Delta T^{-5}$, follows reasonably well the points in the regime III.
}
\label{fig_temps}
\end{figure}

This first crack time $t_c$ has been measured for all our experiments, varying both the impact velocity and the substrate temperature and is shown on figure~\ref{fig_temps}. The open diamonds correspond to the fragmentation time in regime II and closed triangles to the appearance of the first crack in regime III. We observe that in this third regime the experimental points follow reasonably well the $\Delta T^{-5}$ variation predicted by the relation (\ref{tf}) and plotted with a dashed line.
This confirms our model where quasi stationary heat diffusion in the ice layer drives the solidification rate and the first crack appears when the thickness of the ice layer is close to the Griffith length.

Finally, the transition between regimes II and III is expected when $h_0 \sim L_c$. 
Then, the elastic energy in the ice block has to be estimated at the time when the solidification ends, namely when $h(t) = h_0$. Integrating Eq.~\ref{DensityEnergy} on the pancake volume, with the corresponding temperature field (Eq.~\ref{TempProfil}) yields : 
$$ \frac{3E\alpha^2}{2}     \Delta T^2_\text{II-III}  \pi R^2 \int_0^{h_0}   (1-\frac{z}{h_0})^2 dz=\frac{E \alpha^2}{2} \Delta T^2_\text{II-III} \pi R^2 h_0.$$
Balancing this energy with the minimal elastic energy needed to fragment ($ \frac{3}{2} E \alpha^2 \Delta T_\text{I-II}^2 \pi R^2 h_0$), allows us to
obtain the transition temperature $\Delta T_\text{II-III}$ separating the two fracture regimes :
\begin{equation} 
\Delta T_\text{II-III} =\sqrt{3} \Delta T_\text{I-II}.
\label{DeltaT_II-III}
\end{equation}
Taking $\Delta T_\text{I-II} \sim 26^\circ$ computed above leads to $\Delta T_\text{II-III}~\sim~45^\circ$, which is in very good agreement with the experimental transition temperature $T_\text{II-III}^{\text{(exp)}}=42^\circ$.


In conclusion, in this paper the different crack regimes of a frozen water pancake shrunk by cooling and pinned on a non shrinking substrate, are investigated using classical fracture scaling arguments. By increasing the thermal shock, the pancake undergoes two regimes : from fragmentation to hierarchical fracture. The appearance temperature of both regimes are determined, along with the scaling of the first crack time in the hierarchical fracture regime.

\begin{figure}[h]
\center
\includegraphics[width=1.\linewidth]{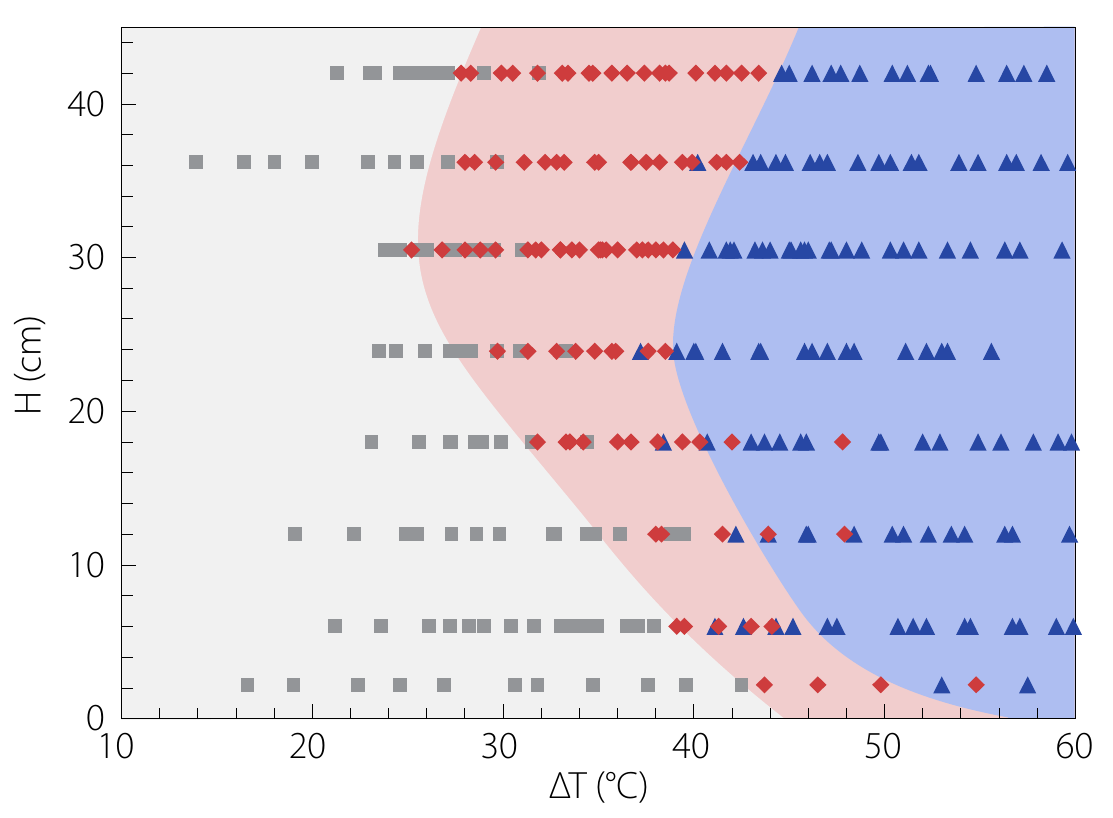}
\caption{Phase diagram for the cracks pattern as the substrate temperature (-$\Delta t$) and the drop impact velocity (here noted by $H$ the height of drop fall) vary. The three regimes observed are represented with the same symbols and same color as on Fig.~\ref{cracks} : white square for regime I, red diamond for regime II and blue triangle for regime III.}
\label{fig_diag}
\end{figure}

Finally, until now only one falling height $H$ has been considered for the drop, which signifies that the shape of the pancake has been kept almost constant. However, drop impact enables a control of the pancake aspect ratio and further on of the cracks patterns of thin structures. Indeed, by varying the impact parameters and the substrate temperature, our experimental set-up allows us to span a large range of spreading dynamics, leading to a broad variety of frozen drop shape~\cite{Yarin06,EFJZ10,JTAR16}.
Figure~\ref{fig_diag} displays the phase diagram as both $H$ and $\Delta T$ vary, where the three main domains of Fig.~\ref{cracks} are retained, proving their universality. However, we observe that the transition temperatures vary non monotonically with the drop falling height: since increasing $H$ decreases the pancake thickness ($h_0$), we would expect the transition temperatures  $\Delta T_\text{I-II} \propto h_0^{-1/2}$ (Eq.~\ref{DeltaT_I-II}) and $ \Delta T_\text{II-III}=\sqrt{3} \Delta T_\text{I-II}$ (Eq.~\ref{DeltaT_II-III}) to increase with $H$, which is only compatible in our experiment for $H$ greater than 25-30 cm. Below this height, the transition temperature decreases and our model becomes wrong, probably because the frozen drop does not have the pancake cylindrical shape of Fig.~\ref{Schema_goutte} anymore.


\bibliographystyle{apsrev4-1}

%


\end{document}